# Brief outbursts in the dwarf nova V1316 Cygni

Jeremy Shears, David Boyd & Gary Poyner

Several brief outbursts were detected in the dwarf nova V1316 Cyg during 2005. These events have an average amplitude of 1.4 magnitude and a duration of <1 to 2 days. Whilst no outburst period could be confirmed, the shortest period between detected events is 10 days. These curious brief outbursts appear to be the normal pattern of behaviour for this system. They are of smaller amplitude and shorter duration than normal outbursts previously reported in this star. V1316 Cyg appears to be a very unusual system and we suggest that further observations by both amateur and professional astronomers could yield important information about the underlying mechanism.

## Introduction

V1316 Cyg was first reported by G. Romano in 1967 and described as an irregular variable.[1] He later revised this classification and it was assigned to the dwarf novae, with a possible outburst period of more than 100 days.[2] Bruch & Schimpke recorded a spectrum of the star during outburst and found it to be typical of an erupting dwarf nova,[3] confirming Romano's final classification. V1316 Cyg is listed in the General Catalogue of Variable Stars[4] as a dwarf nova which is suspected of being a member of the SU UMa family (UGSU star), although there are neither data nor literature references provided to support this suspicion. We have also searched for such supporting data without success. The magnitude range is listed in GCVS as 14.1 to 17.5, i.e. more than 3 magnitudes.

Dwarf novae are known to be interacting binary stars in which a cool main sequence star (the secondary) loses mass to a white dwarf primary. The result is the formation of an accretion disc around the primary. From time to time the accretion disc flips from a cooler, dimmer state, to a hotter, brighter state, resulting in what we see as an outburst in which the star brightens by several magnitudes. Whilst a wide range of outburst behaviour is exhibited in dwarf novae, the smallest outburst amplitude is of the order of two magnitudes[5,6] and the shortest outburst duration is two to three days.[7]

A characteristic of the UGSU subclass of dwarf novae is that they occasionally exhibit 'superoutbursts' which are typically 0.5 to 1 magnitude brighter than normal outbursts and last up to 10 times longer. During a superoutburst, the light curve is characterised by superhumps. These are modulations which are a few percent longer than the orbital period and are thought to be caused by precession of the accretion disc.[8]

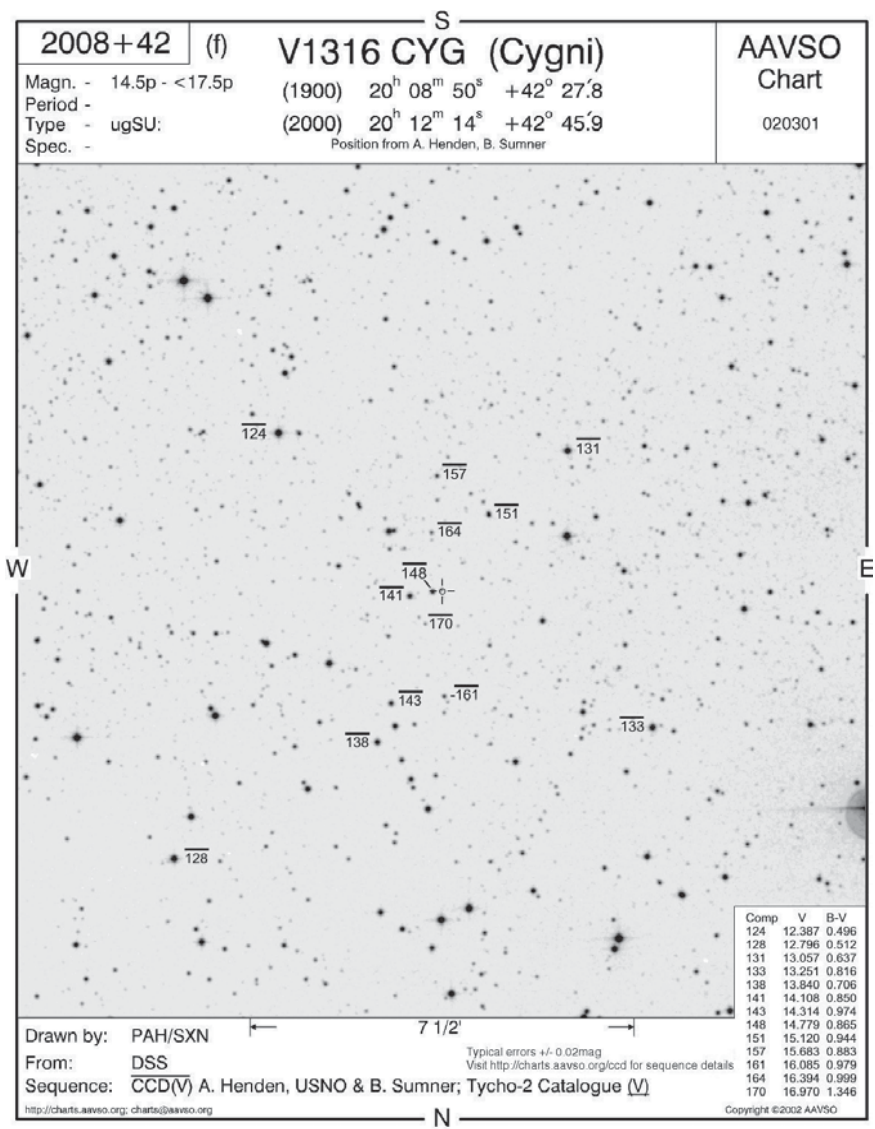

**Figure 1.** AAVSO observing chart for V1316 Cyg. *(Courtesy AAVSO)*

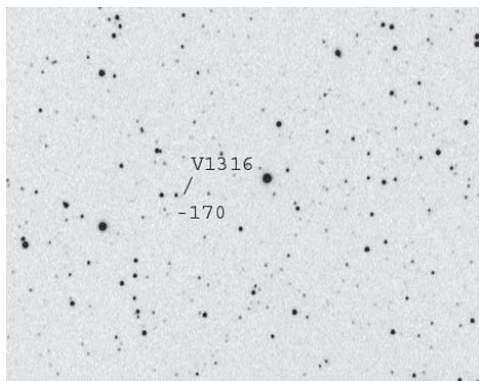
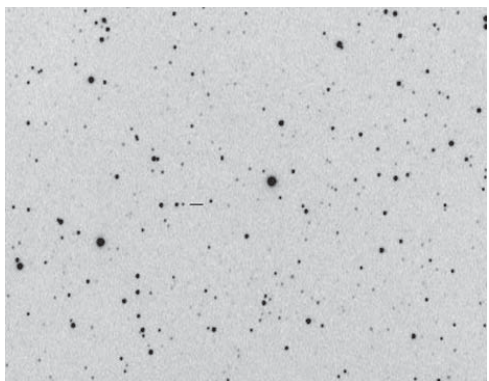

**Figure 2** *(left)*.  V1316 Cyg in quiescence (just visible at about 17.5C), 2005 May 14, 23.34UT. The star just to the west of V1316 Cyg is a 14.8V field star. Also shown is star 170 from the AAVSO chart (16.97V). Field size 10'×8'; N at bottom and E to right.  *(J. Shears.)*

**Figure 3** *(right)*.  V1316 Cyg active at 15.8C, 2005 Sep 5, 19.55 UT.  *(J. Shears.)*

V1316 Cyg remains a poorly studied star and superhumps, which would definitively characterise it as a member of the UGSU family, have yet to be detected. In this paper we describe the results of CCD photometry carried out during the summer of 2005. Our aim is to encourage other observers, both visual and CCD, to include the star in their observing programmes.

## V1316 Cyg and the BAA Recurrent Objects Programme

In an attempt to encourage further observations, Gary Poyner added V1316 Cyg to the Variable Star Section's Recurrent Objects Programme in 1996. The position was taken from the Downes & Shara catalogue[9] and charts were produced and distributed. Initial visual observations showed V1316 Cyg to be varying irregularly by one magnitude between 14.4 and 15.4 during the period from 1996 to 2000, and it was thought that the star had been misclassified as a dwarf nova. However in 2000 September, Bruce Sumner reported on the VSNET and AAVSO online discussion groups that the position given in Downes & Shara was incorrect, and that V1316 Cyg was in fact the faint star (17.58V) some 12 arcsec east of the indicated star. This was deduced from B and V photometry and magnitude variation in quiescence, using photometric data obtained by Henden & Honeycutt for their paper on 'Secondary photometric standards for northern cataclysmic variables and related objects'.[10] Although the previous four years of data on V1316 Cyg had now been rendered incorrect, it had established that the close field star was indeed a small amplitude variable. The online version of the Downes & Shara catalogue now shows the variable correctly identified.[11]

The AAVSO chart for V1316 Cyg (Figure 1) shows the variable and the adjacent field star (labelled '148'). Visually, V1316 Cyg is a challenge to observe due to the proximity of this close field star. The field star also represents a hazard for CCD observers, especially those employing automated reduction software, since false claims of outbursts are still reported which relate to the field star being detected, rather than V1316 Cyg itself.

Only two outbursts have been reported by visual observers in the period 2000 to 2004:

| Date | Mag. | Observer |
|---|---|---|
| 2003 Sep 21 | 14.6 | Mike Simonsen/Dan Taylor |
| 2004 Jun 29 | 15.0 | Mike Simonsen |

## CCD surveillance of V1316 Cyg during 2005

Shears began to monitor the star at the Bunbury Observatory during the summer of 2005. The instrument used is a Takahashi FS102, a 0.1m apochromatic refractor, operating at f/8. The detector is a Starlight Xpress SXV-M7, which is used unfiltered ('C' magnitudes, or clear) with an integration time of 60s. This routinely gives a limiting magnitude of between 17.5C and 18C, although on exceptional nights stars well below mag 18 are detected.

Using this equipment, V1316 Cyg is usually just detectable in quiescence at about mag 17.5C. However, on occasions it is invisible, sometimes due to its being inherently faint, or sometimes due to poor transparency during the observation; these negative results have been omitted from the subsequent analysis. Images of V1316 Cyg in quiescence and whilst active are shown in Figures 2 and 3.

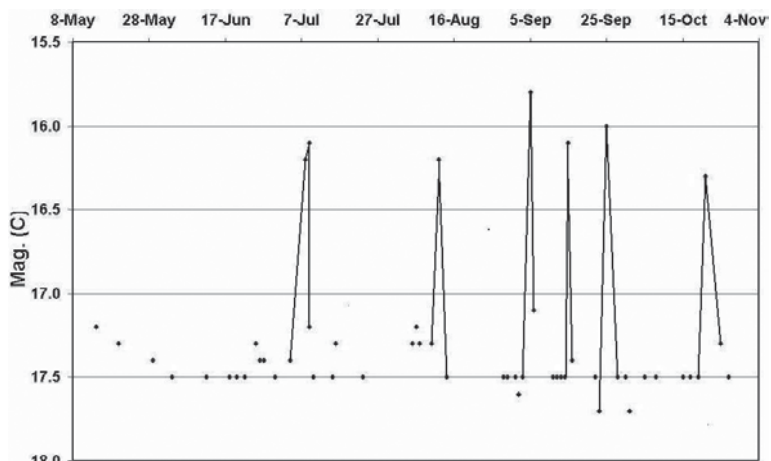

**Figure 4.**  Light curve of V1316 Cyg during 2005, showing six outbursts. Negative observations, i.e. 'less than', have been omitted. *(J. Shears.)*

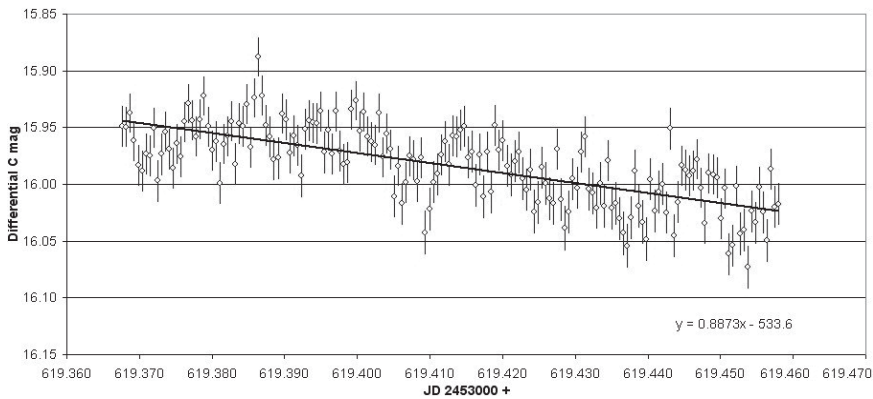

**Figure 5.** Photometry of V1316 Cyg on 2005 September 5. *(D. Boyd.)*

Between 2005 May 17 and October 30, six periods of activity are apparent in which the star brightened significantly to between 15.8C and 16.3C (Figure 4). Table 1 shows the times when each episode was recorded, as well as the observation immediately before and after. Although the data are not sufficiently dense to determine the precise duration of these brightening episodes, what is remarkable about each is that they were relatively short-lived. In three cases the star returned to quiescence within 26h. Incomplete observational coverage of the other outbursts precludes a similar conclusion being drawn about them, although it is evident that in one case quiescence was reached within 48h. Moreover each event was relatively modest in terms of the brightness increase (mean of 1.4 magnitudes) and certainly much less than in the 2003 and 2004 outbursts.

Boyd conducted time resolved photometry, using a 0.35m SCT and an unfiltered Starlight Xpress SXV-H9 CCD camera, on 2005 Sept 5 during one of the outbursts (Figure 5). This showed a mean rate of decline of about 0.9 mag/day with some modulation of about ±0.05 mag. Such a rate of decline is much faster than normal dwarf nova outbursts, which have typical decline rates of 0.3–0.4 mag per day. This again suggests that this event is not a normal outburst.

No specific conclusions can be drawn regarding the outburst period, because of the gaps in the data coverage and the uncertainty in the duration of the outbursts. Furthermore, there are too few events recorded to conclude they are random. The data in Table 1 show that the outburst period, if indeed there is one, could be as short as 10 days. Period analysis of the outburst data using the DFT algorithm in the *Peranso* software package[12] yielded a strong signal at 5 days, but it is clear from Table 1 and Figure 4 that these outbursts do not actually occur every 5 days. No period consistent with all the data can be found.

## Discussion

Our observational data suggest that the multiple outbursts we have detected in V1316 Cyg are not normal dwarf nova outbursts. As discussed in the Introduction, such dwarf nova outbursts are of at least two magnitudes in amplitude and two to three days in duration and in the case of V1316 Cyg the normal outburst amplitude is more than three magnitudes. Since the outbursts we observed are of lesser amplitude and duration, we will describe them as 'brief outbursts'.

What other short-term brightness variations are found in dwarf novae which could explain the brief outbursts in V1316 Cyg? Apparently random variations in brightness are common in dwarf novae.[8] In fact such 'flickering' is considered almost a characteristic feature of these objects.[12] It can occur on many timescales, from a few seconds to hours and can be milli-magnitudes to several tenths of a magnitude in amplitude. Flickering is essentially a continuum of individual flares. Whilst the precise cause or causes of these flares are not fully understood, it is possible that it is because transfer of material from the secondary star to the white dwarf is a turbulent process, not a smooth one. Studies on U Gem suggest that flickering originates from the bright spot where material streaming from the secondary hits the accretion disc in a turbulent manner. By contrast other studies, notably of HT Cas, suggest that the flickering appears to arise predominantly in the turbulent inner regions of the accretion disc.[8] However, the brief outbursts we have observed in V1316 Cyg are discrete events (although there are several of them) and are clearly not part of a continuum of activity.

Short outbursts have been reported in several intermediate polars (IPs), a class of magnetic cataclysmic variable. For example, TV Col has shown outbursts of up to two magnitudes over a period of a few hours,[14] although only 12 outbursts have been recorded since 1934. EX Hya has outbursts of amplitude ~4 magnitude and a typical duration of two days, but some have been observed to last less than one day.[15] Over a 44 year period 15 outbursts have been seen, leading to an outburst period of 1.5 years, corrected for sampling losses. The IP V1223 Sgr has had a single observed outburst of about one magnitude lasting less than one day.[16] It has been suggested that these events are caused by flares on the secondary resulting in a short-lived burst of mass transfer.[14]

**Table 1. Outbursts of V1316 Cyg**

| Date (2005) | Magnitude (C) | Amplitude above mean quiescence (17.5C) | Time from maximum to quiescence |
|---|---|---|---|
| Jul 04.988 | 17.4 | | |
| Jul 09.047 | 16.1 | 1.4 | <22h |
| Jul 09.927 | 17.2 | | |
| | | | |
| Aug 10.906 | 17.3 | | |
| Aug 12.927 | 16.2 | 1.3 | <48h |
| Aug 14.920 | 17.5 | | |
| | | | |
| Sep 03.890 | 17.5 | | |
| Sep 05.830 | 15.8 | 1.7 | <26h |
| Sep 06.877 | 17.1 | | |
| | | | |
| Sep 14.866 | 17.5 | | |
| Sep 15.941 | 16.1 | 1.4 | <22h |
| Sep 16.856 | 17.4 | | |
| | | | |
| Sep 23.883 | 17.7 | | |
| Sep 25.853 | 16.0 | 1.5 | <75h |
| Sep 28.939 | 17.5 | | |
| | | | |
| Oct 19.826 | 17.5 | | |
| Oct 21.813 | 16.3 | 1.2 | <94h |
| Oct 25.785 | 17.3 | | |

Whilst the outbursts in these three IPs are superficially like those seen in V1316 Cyg, the frequency is totally different. Moreover, it appears that V1316 Cyg has two types of outburst: the brief ones which occur every few days and the brighter ones from the historical record, such as in 2003 and 2004, that have a period of more than 100 days. In the three IPs, normal or superoutbursts is the standard pattern of behaviour.

A number of intriguing questions arise from this analysis, which warrant further investigation. Whilst there is no evidence in the literature, could V1316 Cyg itself be an IP? If the brief outbursts are due to mass-transfer events, then does V1316 Cyg have a particularly active secondary or is it prone to the instability that causes these events? However, given the superficial similarity of V1316 Cyg's brief outbursts to events in IPs, is it more likely that the outbursts are due to some interaction between the disc and the magnetosphere? Finally, in TV Col the brief outbursts are seen on top of a permanently superhumping light curve,[15] which presumably implies significant additional mass transfer. Hence, are V1316 Cyg's bright outbursts from the historical record in reality normal disc-instability events, or are they superoutbursts in the UGSU sense?

## Opportunities for further observational study

V1316 Cyg appears to be a unique star which exhibits frequent 'brief outbursts'. Since these outbursts appear to occur rather often, the authors suggest that this star would be a good candidate for further studies, which may reveal more information about the origin of such events in dwarf novae. Moreover, this could be an interesting professional–amateur cooperation project. Amateurs could monitor for these outbursts and alert the professionals so they can carry out multi-wavelength photometry and spectroscopy at high speed. Efforts are currently being made to interest professionals in this project.

V1316 Cyg remains an interesting star to monitor either visually or using CCDs and the authors would like to encourage others to join in. It would be helpful to determine the real frequency and duration of the brief outbursts (it is possible that some have been missed due to incomplete coverage) and to determine whether these are truly random events, or whether there is some periodicity. And do they occur all the time, or has the summer of 2005 been a particularly active time for this star? Identifying future *true* outbursts will also be helpful in refining the outburst period.

Finally, if the star is indeed a member of the UGSU subclass as proposed by the GCVS, superoutbursts are expected. If such a superoutburst is detected, time resolved CCD photometry could be used to look for superhumps, thereby confirming the UGSU classification. Furthermore, since there is an empirical relationship between the superhump period and the orbital period in UGSU stars, the orbital period could be determined.[17]

## Acknowledgments

The authors would like to thank Dr Coel Hellier, University of Keele, for helpful discussions during the preparation of this paper. We are also grateful to Dr Arne Henden, the Director of the American Association of Variable Star Observers, for giving his permission to reproduce the AAVSO chart of V1316 Cyg. Finally, we are indebted to Dr Chris Lloyd and Dr Bill Worraker, for their comments made during the refereeing of this paper and for drawing our attention to supplementary references in the literature.

**Addresses: JS:** 'Pemberton', School Lane, Bunbury, Tarporley, Cheshire CW6 9NR. [bunburyobservatory@hotmail.com]
**DB:** 5 Silver Lane, West Challow, Wantage, Oxon. OX12 9TX. [drsboyd@dsl.pipex.net]
**GP:** 67 Ellerton Road, Kingstanding, Birmingham B44 0QE. [garypoyner@blueyonder.co.uk]

## References

1 Romano G., *IBVS,* **229** (1967)
2 Romano G., *Padova Publ.,* **156** (1969)
3 Bruch A. & Schimpke T., *Astron. Astrophys. Suppl. Ser.*, **93**, 419–439 (1992)
4 *General Catalogue of Variable Stars*, available on line at **http://www.sai.msu.su/groups/cluster/gcvs/**
5 Richter G. & Braeuer H.–J., *Astron. Nachrichten,* **310**, 413–418 (1989)
6 Warner B., *Cataclysmic Variable Stars*, Cambridge University Press, 1995
7 Ak T., Ozkan M. T. & Mattei J. A., *Astron. Astrophys.,* **389**, 478–484 (2002)
8 Hellier C., *Cataclysmic Variable Stars: How and why they vary*, Springer–Verlag, 2001, Ch.10
9 Downes R. & Shara M. M., *Proc. Astron. Soc. Pacific,* **105**, 127 (1993)
10 Henden A. & Honeycutt R. K., *ibid.*, **109**, 441–460 (1997)
11 Downes R. & Shara M. M., *A Catalog and Atlas of Cataclysmic Variables, Living edition*: **http://archive.stsci.edu/prepds/cvcat/index.html**
12 Vanmunster T., *Peranso*, **www.peranso.com**
13 Bruch A., *Astron. Astrophys.,* **266**, 237–265 (1992)
14 Hellier C. & Buckley D. A. H., *MNRAS*, **265**, 766–772 (1993)
15 Hellier C. *et al.*, *ibid.*, **313**, 703–710 (2000)
16 van Amerongen S. & van Paradijs J., *Astron. Astrophys.,* **219**, 195–196 (1989)
17 Stolz B. & Schoembs R., *ibid.,* **132**, 187 (1984)